\documentclass[sigconf,natbib=true,screen]{acmart}
\usepackage[table,dvipsnames]{xcolor} 
\usepackage{subcaption}
\usepackage{multirow}
\usepackage{xcolor}

\AtBeginDocument{%
  }

\setcopyright{acmlicensed}
\copyrightyear{2026}
\acmYear{2026}
\acmDOI{XXXXXXX.XXXXXXX}
\acmConference
{the 49th
International ACM SIGIR Conference on Research and Development in Information Retrieval (SIGIR ’26), July 20–24, 2026, Melbourne, Australia.}{June 03--05,
  2018}{Woodstock, NY}
\acmISBN{978-1-4503-XXXX-X/2018/06}

\begin{document}

\title{Are LLM-Based Retrievers Worth Their Cost? An Empirical Study of Efficiency, Robustness, and Reasoning Overhead
}

%

\author{Abdelrahman Abdallah}
\affiliation{%
  \institution{University of Innsbruck}
  \city{Innsbruck }
  \country{Austria}}
\email{abdelrahman.abdallah@uibk.ac.at}

\author{Jamie Holdcroft}
\affiliation{%
  \institution{UNSW Sydney}
  \city{Sydney}
  \country{Australia}
}
\email{j.holdcroft@student.unsw.edu.au}

\author{Mohammed Ali}
\affiliation{%
  \institution{University of Innsbruck}
  \city{Innsbruck }
  \country{Austria}}
\email{Mohammed.Ali@uibk.ac.at}

\author{Adam Jatowt}
\affiliation{%
  \institution{University of Innsbruck}
  \city{Innsbruck }
  \country{Austria}}
\email{adam.jatowt@uibk.ac.at}





\renewcommand{\shortauthors}{Trovato et al.}

\begin{abstract}
Large language model retrievers improve performance on complex queries, but their practical value depends on efficiency, robustness, and reliable confidence signals in addition to accuracy. We reproduce a reasoning-intensive retrieval benchmark (BRIGHT) across 12 tasks and 14 retrievers, and extend evaluation with cold-start indexing cost, query latency distributions and throughput, corpus scaling, robustness to controlled query perturbations, and confidence use (AUROC) for predicting query success. We also quantify \emph{reasoning overhead} by comparing standard queries to five provided reasoning-augmented variants, measuring accuracy gains relative to added latency. We find that some reasoning-specialized retrievers achieve strong effectiveness while remaining competitive in throughput, whereas several large LLM-based bi-encoders incur substantial latency for modest gains. Reasoning augmentation incurs minimal latency for sub-1B encoders but exhibits diminishing returns for top retrievers and may reduce performance on formal math/code domains. Confidence calibration is consistently weak across model families, indicating that raw retrieval scores are unreliable for downstream routing without additional calibration. We release all code and artifacts for reproducibility. 
\end{abstract}



\maketitle

\section{Introduction}

Neural retrieval has undergone a profound transformation with the advent of Large Language Models (LLMs), which have enabled a new generation of retrieval systems capable of handling complex, reasoning-intensive queries~\cite{su2024bright, abdallah2026mm,gruber2025complextempqa,shao2024reasoni, ali2026recor, abdallah2026tempo}. 
Dense bi-encoders built upon LLM backbones, such as E5-Mistral~\cite{wang2024e5mistral}, Qwen2~\cite{wang2024qwen2}, and specialized reasoning retrievers like ReasonIR~\cite{shao2024reasoni} and Diver~\cite{song2025diver}, have demonstrated remarkable improvements over traditional sparse and dense retrieval methods on benchmarks requiring multi-hop reasoning, compositional understanding, and domain-specific knowledge.
For instance, the BRIGHT benchmark~\cite{su2024bright} was specifically designed to evaluate retrieval under such reasoning-oriented settings, revealing a substantial performance gap: while dense retrievers achieve up to 59.0 nDCG@10 on BEIR~\cite{thakur2021beir}, their effectiveness drops to as low as 18.3 on BRIGHT, a decrease of over 40 absolute points.

However, existing evaluations of LLM-based retrievers, often benchmarked on standard datasets like MS MARCO~\cite{nguyen2016ms}, Natural Questions~\cite{kwiatkowski2019natural}, and HotpotQA~\cite{yang2018hotpotqa}, focus almost exclusively on ranking effectiveness, typically reporting nDCG@10 or Recall@k while abstracting away the substantial computational costs and practical trade-offs that accompany reasoning-intensive retrieval.
This creates a ``hidden costs'' problem in the literature: practitioners reading benchmark results cannot easily determine whether the additional overhead of LLM-based retrievers' larger model sizes, longer inference times, and increased memory requirements is justified by proportional gains in retrieval quality.
The situation is further complicated by the diversity of reasoning-augmented query strategies, where upstream LLMs (GPT-4~\cite{achiam2023gpt}, Claude-3~\cite{TheC3}, Llama-3-70B~\cite{dubey2024llama}, Gemini~\cite{team2024gemini}, GritLM~\cite{muennighoff2024generative}) generate enhanced queries that may improve effectiveness but at the cost of additional latency and API calls.

To address these gaps, we present a comprehensive reproducibility study that goes beyond effectiveness reproduction to characterize the \emph{multi-dimensional trade-offs} of modern retrieval systems. We select the BRIGHT benchmark~\cite{su2024bright} as our evaluation framework because it is the most recent reasoning-intensive retrieval benchmark, comprising 12 diverse tasks that explicitly require multi-step inference, compositional understanding, and domain-specific knowledge capabilities that standard benchmarks such as BEIR and MS~MARCO do not target.
Using the BRIGHT benchmark, we systematically compare a diverse spectrum of retrievers spanning classical sparse retrieval (BM25)~\cite{robertson2009bm25}, traditional dense bi-encoders (SBERT, BGE, Contriever, E5)~\cite{reimers2019sentence,xiao2023bge,izacard2022contriever,wang2022e5}, instruction-tuned models (Instructor-L, Instructor-XL, Qwen, Qwen2)~\cite{su2023one,li2023gte}, and reasoning-specialized retrievers (ReasonIR, RaDeR, Diver)~\cite{shao2024reasoni,das2025rader,song2025diver} under controlled experimental conditions.

Our evaluation extends the original benchmark along dimensions critical for practical deployment: efficiency profiling (indexing time, query latency at p50/p95/p99, throughput)~\cite{wang2023efficiency,mackenzie2024reneuir}, corpus scalability~\cite{pradeep2023generative}, robustness to query perturbations~\cite{penha2020evaluating, arabzadeh2023robustness}, and confidence calibration~\cite{penha2021calibration, cohen2021calibration}.
Notably, we introduce the concept of \emph{Efficiency Ratio}---defined as the ratio of nDCG improvement to latency penalty---to quantify whether reasoning overhead translates proportionally into retrieval quality gains.
Our framework systematically covers five critical dimensions for practical deployment.

 Our main contributions are:
\begin{itemize}
    \item A comprehensive reproducibility study of the BRIGHT benchmark, validating effectiveness results across 12 retrieval tasks and verifying the robustness of reported model rankings.
    \item A systematic efficiency analysis profiling indexing time, query latency distributions, and throughput for sparse, dense, and LLM-based retrievers under controlled conditions.
    \item The introduction of the \emph{Efficiency Ratio} metric to quantify effectiveness--efficiency trade-offs for reasoning-augmented queries generated by five different LLMs (GPT-4, Claude-3, Llama-3-70B, Gemini, GritLM).
    \item An empirical evaluation of retriever robustness under four categories of controlled query perturbations, identifying vulnerability patterns across model families.
    \item A novel analysis of retrieval confidence calibration using AUROC-based measures, characterizing which retrievers produce well-calibrated confidence signals.
\end{itemize}

We believe these findings provide practitioners with the multi-dimensional perspective necessary for informed retriever selection and deployment. 
All experimental code, results, and analysis scripts are publicly released to support transparent and reproducible research.\footnote{\url{https://github.com/DataScienceUIBK/LLM-Retrievers-Beyond-Relevance}}

\section{ Related Work}


Dense retrieval encodes queries and documents into a shared vector space for efficient similarity computation~\cite{karpukhin2020dense}.
Early bi-encoder models such as DPR~\cite{karpukhin2020dense}, Contriever~\cite{izacard2022contriever}, E5~\cite{wang2022e5}, and BGE~\cite{xiao2023bge} progressively improved training strategies---from supervised dual-encoder training to unsupervised contrastive pre-training and web-scale consistency filtering---establishing dense retrieval as the dominant paradigm in neural IR. More recently, dense retrievers have scaled to LLM-sized backbones. Models such as E5-Mistral~\cite{wang2024e5mistral}, GTE~\cite{li2023gte}, and Qwen2~\cite{wang2024qwen2} adapt billion-parameter language models into bi-encoder architectures, achieving state-of-the-art effectiveness on BEIR~\cite{thakur2021beir} and MTEB~\cite{muennighoff2022mteb}. However, this scaling---from BERT's 110M to over 7B parameters~\cite{ma2024finetuning}---raises fundamental questions about the efficiency--effectiveness trade-off, which we directly address in this work. Beyond bi-encoders, late-interaction models (e.g., ColBERT) and sparse neural retrievers (e.g., SPLADE) provide strong first-stage retrieval effectiveness while retaining lexical matching signals~\cite{khattab2020colbert,santhanam2022colbertv2,formal2021splade,formal2022spladev2,gao2021coil}.


Traditional retrieval benchmarks primarily evaluate keyword or semantic matching, leaving complex reasoning demands largely unexamined.
The BRIGHT benchmark~\cite{su2024bright} was introduced to address this gap, comprising 1,398 real-world queries across diverse domains that require multi-step reasoning, compositional understanding, and latent relationship identification.
To meet these demands, reasoning-specialized retrievers have been proposed: ReasonIR~\cite{shao2024reasoni} trains a bi-encoder on synthetic reasoning-intensive queries, while approaches such as Retrieval Augmented Thoughts (RAT)~\cite{wang2024rat}, Aligned Query Expansion (AQE)~\cite{liu2024aqe}, and DEAR~\cite{abdallah2025dear} iteratively revise queries with retrieved information, further integrating reasoning into the retrieval process. Query expansion and feedback have long been central in IR and remain effective in modern pipelines. \cite{rocchio1971relevance,lavrenko2017relevance,bonifacio2022inpars,dai2022promptagator}


Despite substantial effectiveness gains, the computational costs of neural retrieval remain a critical practical concern.
The ReNeuIR workshop series~\cite{mackenzie2024reneuir} has highlighted the efficiency--effectiveness gap in neural IR, emphasizing the need for fast inference and standardized efficiency evaluation protocols.
At the query level, decoupling query and passage encoders can yield up to 12$\times$ latency reduction with minimal quality loss~\cite{cohen2024extremely}, while hybrid sparse--dense systems balance speed with semantic precision~\cite{thakur2021beir}.
At the index level, scaling laws suggest that dense retrieval performance follows power-law scaling with model size and annotation volume~\cite{yao2024scaling}.

\begin{table*}
\centering
\caption{BRIGHT benchmark statistics. Q = number of queries, D = number of documents, D$^+$ = avg positive documents per query. Q Len columns show average query length (GPT-2 tokens) for original queries and CoT-augmented variants.}
\small
\resizebox{0.7\textwidth}{!}{
\begin{tabular}{l|rrr|r|rrrrrr}
\toprule
\multirow{2}{*}{Dataset} & \multicolumn{3}{c|}{Counts} & D Len & \multicolumn{6}{c}{Avg. Query Length (tokens)} \\
& Q & D & D$^+$ & Avg. & Orig. & GPT-4 & Llama-3 & Claude-3 & GritLM & Gemini \\
\midrule
\rowcolor{gray!10}\multicolumn{11}{c}{\textit{StackExchange}} \\
\midrule
Biology            & 103 & 57,359  & 3.6 & 83.6  & 115.2 & 655.6 & 730.2 & 553.0 & 402.2 & 606.0 \\
Earth Science      & 116 & 121,249 & 5.0 & 132.6 & 109.5 & 686.1 & 737.6 & 595.6 & 390.0 & 633.5 \\
Economics          & 103 & 50,220  & 7.8 & 120.2 & 181.5 & 724.3 & 755.2 & 616.3 & 455.3 & 710.5 \\
Psychology         & 101 & 52,835  & 6.9 & 118.2 & 149.6 & 669.3 & 702.1 & 578.5 & 367.5 & 683.5 \\
Robotics           & 101 & 61,961  & 5.1 & 121.0 & 818.9 & 828.0 & 757.9 & 674.1 & 467.3 & 772.6 \\
Stackoverflow      & 117 & 107,081 & 4.1 & 704.7 & 478.3 & 800.1 & 740.4 & 674.4 & 507.9 & 721.6 \\
Sustainable Living & 108 & 60,792  & 5.3 & 107.9 & 148.5 & 723.9 & 764.8 & 610.6 & 396.2 & 722.2 \\
\midrule
\rowcolor{gray!10}\multicolumn{11}{c}{\textit{Coding}} \\
\midrule
Leetcode & 142 & 413,932 & 1.8  & 482.6 & 497.5 & 761.3 & 757.6 & 826.7 & 608.4 & 721.7 \\
Pony     & 112 & 7,894   & 19.8 & 98.3  & 102.6 & 766.8 & 571.0 & 714.3 & 448.8 & 594.3 \\
\midrule
\rowcolor{gray!10}\multicolumn{11}{c}{\textit{Theorems}} \\
\midrule
Aops               & 111 & 188,002 & 4.7 & 250.5 & 117.1 & 910.7 & 716.1 & 634.3 & 672.9 & 715.6 \\
Theoremqa Questions& 194 & 188,002 & 3.2 & 250.5 & 93.4  & 690.5 & 582.1 & 560.6 & 407.6 & 542.8 \\
Theoremqa Theorems & 76  & 23,839  & 2.0 & 354.8 & 91.7  & 720.4 & 629.2 & 578.5 & 454.1 & 552.3 \\
\bottomrule
\end{tabular}}
\label{tab:bright_stats}
\end{table*}

\section{Experimental Setup}
\subsection{Preliminaries}

Given a query $q$ and a corpus $\mathcal{C} = \{d_1, \ldots, d_N\}$, a retrieval system computes a relevance score $s(q, d)$ for each document and returns the top-$k$ ranked results.
We evaluate three paradigms:

\vspace{0.3em}\noindent
\textbf{Sparse retrieval.}
BM25~\cite{robertson2009bm25} scores documents via weighted term matching:
\begin{equation}
    s_{\text{BM25}}(q, d) = \sum_{t \in q} \text{IDF}(t) \cdot \frac{f(t, d) \cdot (k_1 + 1)}{f(t, d) + k_1 \cdot \left(1 - b + b \cdot \tfrac{|d|}{\text{avgdl}}\right)}.
    \label{eq:bm25}
\end{equation}

\vspace{0.3em}\noindent
\textbf{Dense bi-encoder retrieval.}
Separate encoders map queries and documents to dense vectors, and relevance is computed as~\cite{karpukhin2020dense}:
\begin{equation}
    s_{\text{dense}}(q, d) = \text{sim}\!\left(\text{pool}(E_q(q)),\; \text{pool}(E_d(d))\right),
    \label{eq:dense}
\end{equation}
where $\text{sim}(\cdot,\cdot)$ is cosine similarity and $\text{pool}(\cdot)$ denotes mean or \texttt{[CLS]} pooling into $\mathbf{e} \in \mathbb{R}^h$. Models in this family range from BERT-scale (110M) to LLM-scale (7B+).

\vspace{0.3em}\noindent
\textbf{Reasoning-augmented retrieval.}
The BRIGHT benchmark~\cite{su2024bright} provides queries augmented by an upstream LLM $\mathcal{M}$ (e.g., GPT-4, Claude-3):
\begin{equation}
    q' = \mathcal{M}(q, \textit{prompt}), \quad s_{\text{reason}}(q, d) = s(q', d),
    \label{eq:reasoning}
\end{equation}
where the prompt elicits chain-of-thought reasoning to make implicit information needs explicit. This can improve effectiveness but introduces additional query-time latency.

\subsection{ Dataset}
We use the BRIGHT benchmark following its original task definitions, relevance judgments, and evaluation protocol. BRIGHT comprises 12 retrieval tasks spanning diverse domains, with heterogeneous query types that vary in length, structure, and reasoning complexity.

\subsection{Retrieval Models}

The evaluated models fall into the following categories: (1) \textbf{Sparse Retrieval:} We include BM25  as a classical lexical baseline. 
(2) \textbf{Dense Bi-Encoder Models:} We evaluate a range of dense retrievers that encode queries and documents independently into a shared embedding space. This category includes \texttt{sbert}, \texttt{bge}, \texttt{contriever}, \texttt{nomic}, and \texttt{e5}.  
(3) \textbf{Large and Instruction-Tuned Dense Models:} to study the impact of scale and instruction tuning, we include larger dense retrievers such as \texttt{inst-l}, \texttt{inst-xl}, \texttt{qwen}, \texttt{qwen2}, and \texttt{sf}. 
(4) \textbf{Reasoning-Oriented Retrievers:} We further evaluate models explicitly designed for reasoning-intensive retrieval, including \texttt{ReasonIR}, \texttt{RaDeR}, and \texttt{Diver}. 

\begin{table}
\centering
\caption{Retrieval model specifications. Params = trainable parameters, Max Len = maximum context length in tokens. $^\dagger$For SBERT (all-mpnet-base-v2), 384 tokens is commonly recommended, though 512 is the hard cap. 
}
\small
\resizebox{0.35\textwidth}{!}{
\begin{tabular}{lrrl}
\toprule
\textbf{Model} & \textbf{Params} & \textbf{Max Len} & \textbf{Category} \\
\midrule
\rowcolor{gray!10}\multicolumn{4}{c}{\textit{Sparse}} \\
\midrule
BM25 & -- & -- & Lexical \\
\midrule
\rowcolor{gray!10}\multicolumn{4}{c}{\textit{Dense Bi-Encoder}} \\
\midrule
SBERT & 110M & 512$^\dagger$ & Bi-Encoder \\
BGE-Large & 335M & 512 & Bi-Encoder \\
Contriever & 110M & 512 & Bi-Encoder \\
Nomic & 137M & 8{,}192 & Bi-Encoder \\
\midrule
\rowcolor{gray!10}\multicolumn{4}{c}{\textit{LLM-Based Dense}} \\
\midrule
Inst-L & 335M & 512 & Instruction \\
Instr-XL & 1.5B & 512 & Instruction \\
E5-Mistral & 7B & 4{,}096 & LLM Bi-Enc \\
SFR-Mistral  & 7B & 4{,}096 & LLM Bi-Enc \\
GTE-Qwen  & 7B & 32{,}000 & LLM Bi-Enc \\
GTE-Qwen2 & 7B & 32{,}000 & LLM Bi-Enc \\
\midrule
\rowcolor{gray!10}\multicolumn{4}{c}{\textit{Reasoning-Specialized}} \\
\midrule
ReasonIR  & 8B & 131{,}072 & Reasoning \\
RaDeR  & 7B & 32{,}000 & Reasoning \\
Diver-Retriever & 4B & 40{,}000 & Reasoning \\
\bottomrule
\end{tabular}}
\label{tab:model_specs}
\end{table}
\begin{table*}[ht!]
\centering
\caption{Reproduction of BRIGHT effectiveness (nDCG@10). \textbf{Bold} = best within category; \underline{underline} = overall best per task. The \textcolor{gray}{Rpt.} column shows the average reported in the original BRIGHT paper~\cite{su2024bright}; ``--'' indicates models not evaluated in the original study.}
\small
\resizebox{0.65\textwidth}{!}{
\begin{tabular}{l|ccccccc|cc|ccc|c|c}
\toprule
& \multicolumn{7}{c|}{StackExchange} & \multicolumn{2}{c|}{Coding} & \multicolumn{3}{c|}{Theorems} & & \\
Model & Bio. & Earth. & Econ. & Psy. & Rob. & Stack. & Sus. & Leet. & Pony & AoPS & TheoQ. & TheoT. & Avg. & \textcolor{gray}{Rpt.} \\
\midrule
\rowcolor{gray!10}\multicolumn{15}{c}{\textit{Sparse}} \\
\midrule
BM25 & 18.9 & 27.2 & 14.9 & 12.5 & 13.6 & 18.4 & 15.0 & 24.4 & 7.9 & 6.2 & 10.4 & 4.9 & 14.5 & \textcolor{gray}{14.5} \\
\midrule
\rowcolor{gray!10}\multicolumn{15}{c}{\textit{Dense Bi-Encoders (<1B)}} \\
\midrule
BGE & 11.7 & \textbf{24.6} & 16.6 & 17.5 & 11.7 & 10.8 & 13.3 & \textbf{26.7} & 5.7 & 6.0 & 13.0 & 6.9 & 13.7 & \textcolor{gray}{13.7} \\
Instructor-L & 15.2 & 21.2 & 14.7 & 22.3 & 11.4 & \textbf{13.4} & 13.5 & 19.5 & 1.3 & \textbf{8.1} & \textbf{20.9} & 9.1 & 14.2 & \textcolor{gray}{14.2} \\
SBERT & 15.1 & 20.4 & 16.6 & \textbf{22.7} & 8.2 & 11.0 & 15.3 & 26.4 & 7.0 & 5.3 & 20.0 & \textbf{10.8} & \textbf{14.9} & \textcolor{gray}{14.9} \\
Contriever & 9.2 & 13.6 & 10.5 & 12.1 & 9.5 & 9.6 & 8.9 & 24.5 & \underline{\textbf{14.7}} & 7.2 & 10.4 & 3.2 & 11.1 & \textcolor{gray}{--} \\
Nomic & \textbf{16.2} & 19.2 & \textbf{16.9} & 19.0 & \textbf{14.4} & 13.1 & \textbf{15.7} & 25.2 & 4.4 & 5.8 & 13.3 & 6.9 & 14.2 & \textcolor{gray}{--} \\
\midrule
\rowcolor{gray!10}\multicolumn{15}{c}{\textit{LLM-Based Dense (>1B)}} \\
\midrule
E5-Mistral & 18.6 & 26.0 & 15.5 & 15.8 & 16.3 & 11.2 & 18.1 & 28.7 & 4.9 & 7.1 & 26.1 & 26.8 & 17.9 & \textcolor{gray}{17.9} \\
SFR-Mistral & 19.1 & 26.7 & 17.8 & 19.0 & 16.3 & 14.4 & 19.1 & 27.4 & 2.0 & 7.4 & 24.3 & 26.0 & 18.3 & \textcolor{gray}{18.3} \\
Instructor-XL & 21.6 & 34.3 & \underline{\textbf{22.4}} & \textbf{27.4} & \textbf{18.2} & 21.7 & 19.1 & 27.5 & 5.0 & 8.5 & 15.6 & 5.9 & 18.9 & \textcolor{gray}{18.9} \\
GTE-Qwen1.5 & 30.6 & 36.4 & 17.8 & 24.6 & 13.1 & \textbf{22.2} & 14.8 & 25.5 & \textbf{9.9} & 14.4 & 27.8 & 32.9 & 22.5 & \textcolor{gray}{22.5} \\
GTE-Qwen2 & \textbf{31.8} & \textbf{40.7} & 16.2 & 26.6 & 12.5 & 15.9 & \textbf{20.7} & \textbf{31.1} & 1.3 & \underline{\textbf{15.1}} & \textbf{32.3} & \textbf{35.5} & \textbf{23.3} & \textcolor{gray}{--} \\
\midrule
\rowcolor{gray!10}\multicolumn{15}{c}{\textit{Reasoning-Specialized}} \\
\midrule
ReasonIR & 25.4 & 27.8 & 20.3 & 29.7 & 19.0 & \underline{\textbf{23.7}} & 21.6 & 33.2 & 12.8 & \textbf{14.7} & 34.0 & 27.1 & 24.1 & \textcolor{gray}{--} \\
RaDeR & 23.4 & 26.1 & 17.3 & 25.5 & 14.2 & 21.3 & 17.0 & 31.6 & 12.6 & 12.7 & \underline{\textbf{42.4}} & \underline{\textbf{38.1}} & 23.5 & \textcolor{gray}{--} \\
Diver & \underline{\textbf{42.0}} & \underline{\textbf{46.6}} & \textbf{21.8} & \underline{\textbf{35.1}} & \underline{\textbf{21.5}} & \underline{\textbf{23.7}} & \underline{\textbf{25.5}} & \underline{\textbf{37.4}} & \textbf{13.4} & 10.6 & 38.2 & 37.1 & \underline{\textbf{29.4}} & \textcolor{gray}{--} \\
\bottomrule
\end{tabular}}
\label{tab:reproduction}
\end{table*}
\subsection{Evaluation Metrics}
We evaluate retrievers along four dimensions. \textbf{Effectiveness} is measured by nDCG@10. \textbf{Efficiency} includes indexing time, peak index storage, and query-time latency at the 50th, 95th, and 99th percentiles (p50/p95/p99) with throughput in queries per second (QPS). To assess the cost--benefit of reasoning-augmented queries, we report an \emph{Efficiency Ratio} (nDCG@10 gain per added query latency). 
\begin{equation}
    ER = \frac{\Delta \text{nDCG@10}}{\Delta \text{Latency (ms)}}.
    \label{eq:efficiency_ratio}
\end{equation}
\textbf{Robustness} is measured as nDCG@10 drop under controlled perturbations (paraphrasing, synonyms, adversarial insertion, length changes). \textbf{Reliability} is evaluated by AUROC using the top-1 retrieval score as a confidence signal.

\subsection{Experimental Setup}

All experiments are conducted on a compute node equipped with 4$\times$ NVIDIA H100-80GB GPUs, 512GB system memory, and AMD EPYC processors. We use CUDA 12.4 with cuDNN 9.1, Torch 2.8.0 for model inference, and the HuggingFace Transformers library (v4.57) for tokenization and model loading. All dense models use FP16 precision and official HuggingFace checkpoints using Rankify~\cite{abdallah2025rankify}. We fix random seeds across all experiments and report timing measurements as the median of three independent runs.

\section{Reproducibility Results}
\label{sec:reproducibility}

\noindent
\textbf{RQ1: Can the BRIGHT benchmark results be reproduced, and what are the true computational costs of each retriever?}

We follow the official BRIGHT benchmark implementation\footnote{\url{https://github.com/xlang-ai/BRIGHT}} and reproduce the evaluation across all 12 tasks using the official queries, corpora, and relevance judgments.
We evaluate 14 retrievers (Table~\ref{tab:model_specs}) using official HuggingFace checkpoints with FP16 inference. 
All document corpora are re-indexed from scratch (cold-start, no pre-existing caches).

\subsection{Effectiveness Reproduction}

Table~\ref{tab:reproduction} presents our reproduced results alongside the originally reported averages (gray \textcolor{gray}{Rpt.} column).
For all eight models that overlap with the original BRIGHT evaluation, our results match the reported values exactly or within rounding: BM25 (14.5), BGE (13.7), SBERT (14.9), E5-Mistral (17.9), SFR-Mistral (18.3), Instructor-XL (18.9), and GTE-Qwen1.5 (22.5) all agree precisely. Per-task discrepancies are consistently below 0.5 nDCG@10, with the minor exceptions attributable to non-determinism in mixed-precision inference for Qwen variants. Beyond reproduction, we extend the evaluation with six models not included in the original study: Contriever, Nomic, GTE-Qwen2, ReasonIR, RaDeR, and Diver-Retriever.
The relative ranking of model families is consistent and robust: reasoning-specialized retrievers substantially outperform all other categories, with Diver-Retriever achieving 29.4 nDCG@10---a 6.1-point gain over the best LLM-based dense model (GTE-Qwen2 at 23.3) and a 14.5-point gain over the best sub-1B model (SBERT at 14.9).

\subsection{Computational Cost of Reproduction}

The original BRIGHT benchmark does not report indexing times or computational costs.
To complete the reproducibility picture, we measure cold-start indexing for all 14 models, reporting total time and throughput (Table~\ref{tab:indexing}).
For dense retrievers, indexing refers to computing and caching document embeddings (no ANN index such as FAISS/HNSW/IVF is built). 
Query latency is measured with exact retrieval via a full cosine-similarity scan over all document embeddings.  We use a batch size of 1 for both document and query encoding.
\begin{table}
\centering
\caption{Cold-start indexing efficiency. Times summed across all 12 BRIGHT tasks.
Peak GPU memory reports the maximum PyTorch CUDA memory allocated per GPU during indexing; models loaded with sharded across 4 GPUs, so values are per-GPU  and exclude non-PyTorch allocations.}
\small
 \resizebox{0.40\textwidth}{!}{
\begin{tabular}{lrrrr}
\toprule
\textbf{Model} & \textbf{Total Time (s)} & \textbf{Avg. Docs/sec} & \textbf{Peak GPU Mem (GB)} \\
\midrule
\rowcolor{gray!10}\multicolumn{4}{c}{\textit{Sparse}} \\
\midrule
BM25 & 219 & 7,950.3 & -- \\
\midrule
\rowcolor{gray!10}\multicolumn{4}{c}{\textit{Dense (<1B)}} \\
\midrule
SBERT & 2,055 & 937.0 & 0.41 \\
Inst-L & 27,564 & 84.9 & 1.25 \\
BGE & 5,616 & 347.3 & 1.25 \\
Contriever & 7,235 & 195.8 & 0.41 \\
Nomic & 14,532 & 97.2 & 0.51 \\
\midrule
\rowcolor{gray!10}\multicolumn{4}{c}{\textit{Dense (>1B)}} \\
\midrule
Inst-XL & 82,667 & 28.7 & 4.63 \\
E5-Mistral & 144,914 & 13.1 & 3.09 \\
SFR-Mistral & 144,923 & 13.1 & 3.09 \\
GTE-Qwen & 116,799 & 15.4 & 3.51 \\
GTE-Qwen2 & 115,971 & 15.6 & 3.50 \\
\midrule
\rowcolor{gray!10}\multicolumn{4}{c}{\textit{Reasoning-Specialized}} \\
\midrule
ReasonIR & 40,304 & 36.9 & 13.98 \\
RaDeR & 39,503 & 37.4 & 13.21 \\
Diver & 21,784 & 68.2 & 15.09 \\
\bottomrule
\end{tabular} }
\label{tab:indexing}
\end{table}

BM25 completes indexing in 219s (7,950 docs/sec).
Among sub-1B dense models, SBERT is fastest (937 docs/sec), while Instructor-L drops to 85 docs/sec.
LLM-based dense models are substantially slower: E5-/SFR-Mistral index at just 13 docs/sec, and GTE-Qwen1.5/2 at 15--16 docs/sec.
Notably, reasoning-specialized models are \emph{faster} than general LLM-based dense models (37--68 docs/sec vs.\ 13--16 docs/sec) despite comparable or better effectiveness.
While indexing is a one-time cost, these differences become critical when corpora are frequently updated or re-indexed.
Query-time inference latency, which more directly impacts user-facing performance, is analyzed in detail in \S\ref{sec:efficiency}.
\subsection{Lessons Learned}

Three findings from our reproduction merit explicit discussion.
First, the \emph{model ranking reported by BRIGHT is robust}: the order of sparse $<$ dense $<$ instruction-tuned $<$ reasoning-specialized holds is consistent under our independent setup, confirming that these effectiveness gaps are not artifacts of specific hardware or software configurations.
Second, \emph{effectiveness comparisons without cost context are incomplete}: the 3.4-point nDCG@10 gain of E5-Mistral over BM25 obscures a 662$\times$ indexing slowdown, and the 6.1-point gap between Diver and GTE-Qwen2 comes at substantially lower latency for Diver (27.6ms vs.\ 209.3ms, see Table~\ref{tab:latency}).
Third, the original BRIGHT evaluation omits several strong baselines---GTE-Qwen2 (23.3) and the reasoning-specialized family (ReasonIR 24.1, RaDeR 23.5, Diver 29.4)---that substantially shift the state-of-the-art picture on this benchmark.
These observations motivate the extended analyses in the following sections.

\begin{table}
\centering
\caption{Query latency and throughput. Latency percentiles measured over all 12 BRIGHT tasks with batch size 1 and cached document embeddings. 
}
\small
\resizebox{0.40\textwidth}{!}{
\begin{tabular}{lrrrrrr}
\toprule
\textbf{Model} & \textbf{QPS} & \textbf{Mean (ms)} & \textbf{p50} & \textbf{p95} & \textbf{p99} & \textbf{nDCG@10} \\
\midrule
\rowcolor{gray!10}\multicolumn{7}{c}{\textit{Sparse}} \\
\midrule
BM25 & 54.4 & 61.7 & 61.6 & 67.0 & 70.8 & 14.5 \\
\midrule
\rowcolor{gray!10}\multicolumn{7}{c}{\textit{Dense (<1B)}} \\
\midrule
SBERT & 95.0 & 11.9 & 11.9 & 11.9 & 11.9 & 14.9 \\
BGE & 58.3 & 19.0 & 19.0 & 19.0 & 19.0 & 13.7 \\
Contriever & 84.9 & 12.6 & 12.6 & 12.6 & 12.6 & 11.1 \\
Nomic & 64.9 & 16.5 & 16.5 & 16.5 & 16.5 & 14.2 \\
Inst-L & 59.5 & 22.1 & 22.1 & 22.1 & 22.1 & 14.2 \\
\midrule
\rowcolor{gray!10}\multicolumn{7}{c}{\textit{LLM-Based Dense (>1B)}} \\
\midrule
Inst-XL & 26.2 & 53.4 & 53.4 & 53.4 & 53.4 & 18.9 \\
E5-Mistral & 5.5 & 230.8 & 177.2 & 565.6 & 706.1 & 17.9 \\
SFR-Mistral & 5.1 & 239.9 & 184.9 & 585.8 & 751.5 & 18.3 \\
GTE-Qwen & 5.7 & 215.5 & 169.1 & 472.1 & 866.9 & 22.5 \\
GTE-Qwen2 & 6.0 & 209.3 & 164.4 & 455.5 & 844.0 & 23.3 \\
\midrule
\rowcolor{gray!10}\multicolumn{7}{c}{\textit{Reasoning-Specialized}} \\
\midrule
ReasonIR & 35.5 & 37.9 & 37.9 & 37.9 & 37.9 & 24.1 \\
RaDeR & 28.0 & 37.4 & 30.2 & 57.7 & 104.3 & 23.5 \\
Diver & 47.3 & 27.6 & 27.6 & 27.6 & 27.6 & 29.4 \\
\bottomrule
\end{tabular}}

\vspace{1mm}
\label{tab:latency}
\end{table}
\section{Efficiency and Reasoning Overhead}
\label{sec:efficiency}

Having confirmed reproducibility and established the computational cost of indexing (\S\ref{sec:reproducibility}), we now turn to query-time efficiency and reasoning overhead---dimensions absent from the original BRIGHT evaluation:

\noindent
\textbf{RQ2: What are the efficiency--effectiveness trade-offs across retriever families?}

\noindent
\textbf{RQ3: Is the effectiveness gain from reasoning-augmented queries worth the additional latency, and when does it help?}


\subsection{Query-Time Efficiency}

Table~\ref{tab:latency} reports throughput, mean latency, and tail latencies using cached document embeddings to isolate query-encoding cost from indexing overhead.
The results reveal efficiency differences spanning roughly two orders of magnitude across model families, a dimension entirely absent from the original BRIGHT evaluation. Among sub-1B bi-encoders, SBERT is fastest (95.0~QPS; 11.9\,ms mean), followed by Contriever (84.9~QPS) and Nomic (64.9~QPS), while Inst-L reaches 59.5~QPS (22.1\,ms).
These models exhibit near-deterministic latency at batch size~1, with identical percentile values because their short query encoding completes within a tight time window regardless of query content.
LLM-based dense models are dramatically slower: E5-Mistral and SFR-Mistral average only 5.1--5.5~QPS ($\sim$230--240\,ms) with heavy tail latency (p99 $>$ 700\,ms), driven by variable-length query tokenization and larger transformer forward passes.
GTE-Qwen/Qwen2 are marginally faster (5.7--6.0~QPS) but still exhibit p99 $>$ 840\,ms, making them unsuitable for latency-sensitive applications.

Reasoning-specialized retrievers emerge as a surprisingly efficient family.
Despite having 4--8B parameters, Diver reaches 47.3~QPS (27.6\,ms)---comparable to sub-1B models---while achieving the highest effectiveness on BRIGHT (29.4 nDCG@10).
ReasonIR and RaDeR also outperform LLM-based dense models in throughput (28--35 QPS vs.\ 5--6 QPS) at higher or comparable effectiveness.
This efficiency advantage likely stems from architectural optimizations and shorter effective sequence lengths compared to general-purpose 7B encoders.
BM25 provides 54.4~QPS on CPU alone with the most stable tail latency (p99~=~70.8\,ms), remaining a strong baseline when no GPU is available.
\begin{figure}[t]
\centering
\includegraphics[width=\columnwidth]{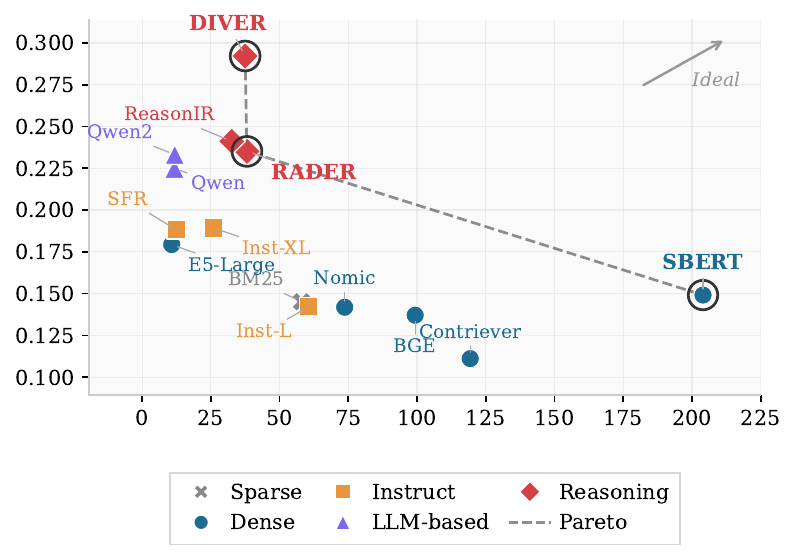}
\caption{Pareto frontier: nDCG@10 vs.\ throughput (QPS). Points above and to the right dominate; dashed line connects Pareto-optimal models.}
\label{fig:pareto}
\end{figure}



\subsection{Pareto Frontier}

Figure~\ref{fig:pareto} plots nDCG@10 against throughput to identify Pareto-optimal retrievers---models for which no other model achieves both higher throughput and higher effectiveness.
The frontier is formed by three models: \textbf{SBERT} (95~QPS, 14.9 nDCG@10) in the high-throughput regime, \textbf{RaDeR} (28~QPS, 23.5) in the balanced region, and \textbf{Diver} (47.3~QPS, 29.4) achieving the strongest effectiveness at competitive throughput.

The most striking finding is that LLM-based dense models (E5-Mistral, SFR-Mistral, GTE-Qwen, GTE-Qwen2) are \emph{all Pareto-dominated}: reasoning-specialized models achieve equal or higher effectiveness at 5--9$\times$ higher throughput.
For example, Diver (29.4 nDCG@10, 47.3~QPS) dominates GTE-Qwen2 (23.3, 6.0~QPS) on both axes simultaneously.
This challenges the implicit assumption that larger encoders justify their cost through proportionally higher effectiveness.
In practice, model selection depends on deployment constraints: SBERT for strict latency budgets ($<$15\,ms per query), Diver when moderate latency ($\sim$28\,ms) is acceptable and effectiveness is prioritized, and BM25 as a no-GPU fallback.


\begin{table}
\centering
\caption{Reasoning query effectiveness (nDCG@10, 12-task averages). \textbf{Orig.}\ = standard BRIGHT queries; other columns use reasoning-augmented variants generated by each LLM.}
\small
\resizebox{0.45\textwidth}{!}{
\begin{tabular}{lcccccc|c}
\toprule
\textbf{Model} & \textbf{Orig.} & \textbf{GPT-4} & \textbf{Llama-3} & \textbf{Claude-3} & \textbf{GritLM} & \textbf{Gemini} & \textbf{Avg.} \\
\midrule
\rowcolor{gray!10}\multicolumn{8}{c}{\textit{Sparse}} \\
\midrule
BM25            & 14.5 & 27.0 & 25.7 & 26.8 & 19.4 & 23.9 & 22.9 \\
\midrule
\rowcolor{gray!10}\multicolumn{8}{c}{\textit{Dense (<1B)}} \\
\midrule
SBERT           & 14.9 & 17.7 & 16.3 & 16.4 & 13.9 & 15.5 & 15.8 \\
BGE             & 13.7 & 22.0 & 20.7 & 21.1 & 16.0 & 18.7 & 18.7 \\
Contriever      & 11.1 & 22.5 & 20.6 & 21.5 & 15.2 & 19.6 & 18.4 \\
Nomic           & 14.2 & 21.3 & 20.1 & 20.2 & 14.9 & 18.3 & 18.2 \\
Inst-L          & 14.2 & 23.5 & 22.7 & 22.1 & 15.8 & 20.8 & 19.9 \\
\midrule
\rowcolor{gray!10}\multicolumn{8}{c}{\textit{LLM-Based Dense (>1B)}} \\
\midrule
Inst-XL         & 18.9 & 26.9 & 26.3 & 26.4 & 22.4 & 24.5 & 24.2 \\
E5-Mistral      & 17.9 & 22.2 & 19.9 & 21.4 & 17.6 & 19.6 & 19.8 \\
SFR-Mistral     & 18.3 & 22.0 & 20.0 & 21.7 & 17.4 & 20.1 & 19.9 \\
GTE-Qwen        & 22.5 & 24.7 & 23.3 & 24.7 & 19.8 & 22.6 & 22.9 \\
GTE-Qwen2       & 23.3 & 26.2 & 24.0 & 24.6 & 20.3 & 22.8 & 23.5 \\
\midrule
\rowcolor{gray!10}\multicolumn{5}{c}{\textit{Reasoning-Specialized}} \\
\midrule
ReasonIR        & 24.1 & 30.2 & 30.0 & 29.5 & 24.1 & 27.9 & 27.6 \\
RaDeR           & 23.5 & 26.8 & 26.0 & 25.6 & 21.1 & 25.0 & 24.7 \\
Diver           & 29.4 & 32.3 & 32.3 & 32.9 & 27.6 & 30.9 & 30.9 \\
\midrule
\textbf{Avg.}   & 18.6 & 24.7 & 23.4 & 23.9 & 18.9 & 22.2 & 22.0 \\
\bottomrule
\end{tabular}}
\label{tab:reasoning_gain}
\end{table}
\subsection{Reasoning Augmentation: Gains and Cost--Benefit}

The BRIGHT benchmark provides reasoning-augmented query variants generated by five LLMs (GPT-4, Llama-3-70B, Claude-3, Gemini, GritLM).
These queries are 4--8$\times$ longer than originals (Table~\ref{tab:bright_stats}), incorporating chain-of-thought reasoning to surface implicit information needs.
Table~\ref{tab:reasoning_gain} reports effectiveness across all five augmentation sources, and Table~\ref{tab:reasoning_efficiency} analyzes the cost--benefit trade-off for GPT-4, the strongest-performing variant.

\vspace{0.3em}\noindent
\textbf{Effectiveness gains.}
Across augmentation sources, GPT-4 produces the largest average improvement (+6.1 nDCG@10 over original queries, from 18.6 to 24.7), followed by Claude-3 (+5.3) and Llama-3 (+4.8).
Gemini yields moderate gains (+3.6), while GritLM is the weakest source (+0.3 on average) and occasionally degrades individual model performance below the original queries.
The magnitude of improvement is inversely related to base model strength.
Weaker retrievers gain the most: BM25 improves from 14.5 to 27.0 with GPT-4 (+12.5 points), and Contriever from 11.1 to 22.5 (+11.4, a 103\% relative increase).
Stronger models benefit less in absolute terms: Diver improves from 29.4 to 32.3 (+2.9), and GTE-Qwen2 from 23.3 to 26.2 (+2.9).
However, reasoning-specialized retrievers still benefit meaningfully (ReasonIR: 24.1~$\rightarrow$~30.2, +6.1), suggesting that reasoning queries and reasoning-trained encoders capture \emph{complementary} aspects of query understanding---the augmented queries supply explicit chain-of-thought context that even specialized training does not fully internalize.

\begin{figure}[t!]
\centering
\centering
\includegraphics[width=0.45\textwidth]{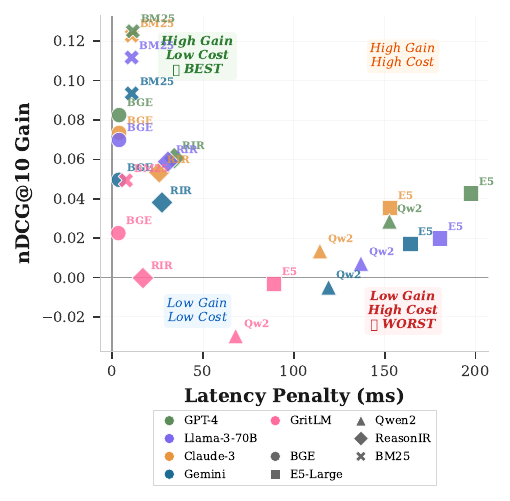}
\caption{nDCG@10 gain vs.\ latency penalty per model. Upper-left quadrant (high gain, low cost) is ideal. 
}
\label{fig:reasoning_quadrant}
\end{figure}

\vspace{0.3em}\noindent
\textbf{Cost--benefit analysis.}
For sub-1B bi-encoders, reasoning queries add negligible latency (within measurement noise, marked $^\dagger$ in Table~\ref{tab:reasoning_efficiency}) because these models process even long queries in under 20\,ms.
This makes reasoning augmentation essentially ``free'' for lightweight models, yielding gains of +2.8 to +11.4 nDCG@10 at zero practical cost.
BM25 shows the best measurable trade-off (ER~=~0.90: +12.5 points for only 13.8\,ms added), and Diver is also highly efficient (ER~=~0.45: +2.9 points for 6.4\,ms).
In contrast, LLM-based dense models exhibit uniformly poor ratios (ER~$\leq$~0.07): E5-Mistral gains only +4.3 points while adding 170.8\,ms per query, and GTE-Qwen2 gains +2.9 for 123.8\,ms---penalties that would be prohibitive in production systems handling thousands of queries per second.
Figure~\ref{fig:reasoning_quadrant} confirms this pattern visually: sub-1B models and Diver cluster in the desirable upper-left quadrant (high gain, low cost), while LLM-based dense models occupy the unfavorable lower-right region.

\begin{table}
\centering
\caption{Efficiency Ratio (ER) for GPT-4 reasoning augmentation. Gain = absolute nDCG@10 improvement. Penalty = added mean query latency (ms). Verdict: \emph{Highly Eff.}\ (ER $\geq$ 0.3), \emph{Worth It} (0.1 $\leq$ ER $<$ 0.3), \emph{Not Worth It} (ER $<$ 0.1). $^\dagger$Penalty within measurement noise ($<$3\,ms); ER not meaningful.}
\small
\resizebox{0.35\textwidth}{!}{
\begin{tabular}{lcccc}
\toprule
\textbf{Model} & \textbf{Gain} & \textbf{Penalty (ms)} & \textbf{ER} & \textbf{Verdict} \\
\midrule
\rowcolor{gray!10}\multicolumn{5}{c}{\textit{Sparse}} \\
\midrule
BM25 & +12.5 & 13.8 & 0.90 & Highly Eff. \\
\midrule
\rowcolor{gray!10}\multicolumn{5}{c}{\textit{Dense (<1B)}} \\
\midrule
SBERT$^\dagger$ & +2.8 & $\approx$0 & -- & Free Gain \\
BGE$^\dagger$ & +8.3 & $\approx$0 & -- & Free Gain \\
Contriever$^\dagger$ & +11.4 & $\approx$0 & -- & Free Gain \\
Nomic$^\dagger$ & +7.1 & $\approx$0 & -- & Free Gain \\
Inst-L & +9.3 & 42.3 & 0.22 & Worth It \\
\midrule
\rowcolor{gray!10}\multicolumn{5}{c}{\textit{LLM-Based Dense (>1B)}} \\
\midrule
Inst-XL & +8.0 & 125.5 & 0.06 & Not Worth It \\
E5-Mistral & +4.3 & 170.8 & 0.02 & Not Worth It \\
SFR-Mistral & +3.7 & 55.8 & 0.07 & Not Worth It \\
GTE-Qwen & +2.2 & 129.3 & 0.02 & Not Worth It \\
GTE-Qwen2 & +2.9 & 123.8 & 0.02 & Not Worth It \\
\midrule
\rowcolor{gray!10}\multicolumn{5}{c}{\textit{Reasoning-Specialized}} \\
\midrule
ReasonIR & +6.1 & 34.2 & 0.18 & Worth It \\
RaDeR & +3.3 & 29.8 & 0.11 & Worth It \\
Diver & +2.9 & 6.4 & 0.45 & Highly Eff. \\
\bottomrule
\end{tabular}}
\label{tab:reasoning_efficiency}
\end{table}

\subsection{Task-Specific Reasoning Effects}

\begin{figure}[t!]
\centering
\centering
\includegraphics[width=0.45\textwidth]{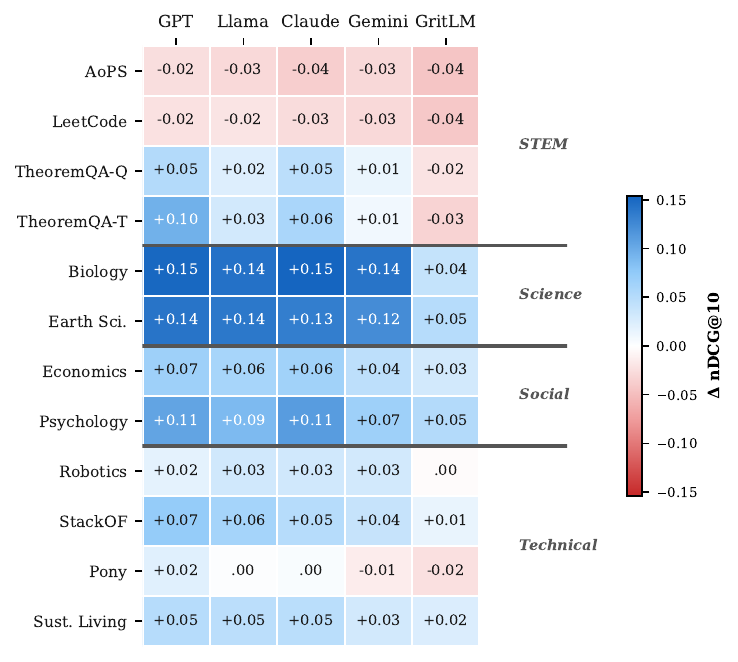}
\caption{Task-level nDCG@10 gain (averaged across retrievers). 
}
\label{fig:reasoning_heatmap}
\end{figure}

Figure~\ref{fig:reasoning_heatmap} reveals that reasoning gains are strongly task-dependent, a finding with direct implications for deployment.
Science and social StackExchange domains benefit the most: Biology gains +0.15 nDCG@10 with GPT-4 and Claude-3 augmentation, Earth Science +0.14, and Psychology +0.11.
These domains feature queries with complex, multi-step information needs where chain-of-thought reasoning can explicitly unpack latent requirements that a short query fails to convey.
Economics and StackOverflow show moderate gains (+0.05 to +0.07).

In contrast, AoPS and LeetCode consistently \emph{degrade} under all five reasoning variants ($-$0.02 to $-$0.04 nDCG@10), and Pony shows near-zero effect.
For AoPS and LeetCode, queries are already precise mathematical or code specifications; the added reasoning verbosity introduces noise (irrelevant tokens, tangential explanations) that dilutes the signal for embedding-based matching.
GritLM consistently yields the smallest gains across tasks and is the only augmentation source that degrades average performance, suggesting that smaller LLMs may lack the reasoning depth needed to produce useful query augmentations.

These results argue strongly against universal deployment of reasoning augmentation.
Instead, a task-aware routing strategy---applying augmentation selectively based on domain characteristics---would maximize gains while avoiding degradation on formal/structured query types.
\subsection{Long-Context Effects}

We compare BRIGHT's standard short-context setting against its long-document evaluation (unsplit web pages) on the eight tasks (Biology, Earth Science, Economics, Psychology, StackOverflow, Sustainable Living, Pony ) where both are available (Table~\ref{tab:long_context}). The short-context setting chunks documents to fit within each model's maximum input length, while the long-context setting preserves full web pages, reducing the corpus size but retaining cross-paragraph evidence.

Switching to long documents substantially improves all models, but the magnitude depends heavily on context window capacity.
LLM-based dense models show the largest gains, as they can now exploit their 4K--32K context windows: E5-Mistral jumps from 15.8 to 44.2 nDCG@10 (+28.4 points); SFR-Mistral from 16.8 to 46.7 (+29.9); and GTE-Qwen/Qwen2 reach 47.9/47.5.
Recall@10 improvements are even more dramatic, with E5-Mistral rising from 18.6 to 64.6 and SFR-Mistral from 20.9 to 67.6.
Diver achieves the best long-context nDCG@10 (48.2) and Recall@10 (69.6), confirming its strong performance across settings. Crucially, models with short context windows benefit far less.
Nomic (8K max length but only 137M parameters) improves modestly (14.9~$\rightarrow$~18.1 nDCG@10), and sub-512-token models like SBERT and BGE see intermediate gains.
This confirms that \emph{short-context chunking is a major bottleneck} in the standard BRIGHT evaluation: when evidence is distributed across long documents, models that truncate input lose critical passages.
The practical implication is that long-context capacity should be weighted heavily in retriever selection for domains with lengthy source documents.

\section{Robustness, Hybrid Strategies, and Reliability}
\label{sec:robustness}

We now evaluate whether retrieval systems are stable under realistic query variations, whether hybrid fusion can improve weaker models, and whether retriever confidence scores are useful for downstream decision-making.

\noindent
\textbf{RQ4: How robust are retrievers to query perturbations, can hybrid fusion bridge effectiveness gaps, and do retrieval scores provide reliable confidence estimates?}
\begin{table}[t]
\centering
\caption{Short-context vs.\ long-context retrieval on eight BRIGHT long-document tasks. Long-context uses unsplit web pages (reduced search space). Metrics are averaged over the eight tasks. \colorbox{green!10}{Light} to \colorbox{green!50}{dark} green shading indicates the magnitude of improvement from short to long context.}
\resizebox{0.48\textwidth}{!}{
\begin{tabular}{lcccc|cccc}
\toprule
& \multicolumn{4}{c|}{\textbf{Short-Ctx}} & \multicolumn{4}{c}{\textbf{Long-Ctx}} \\
\textbf{Model} & \textbf{nDCG@5} & \textbf{nDCG@10} & \textbf{R@5} & \textbf{R@10} &
                \textbf{nDCG@5} & \textbf{nDCG@10} & \textbf{R@5} & \textbf{R@10} \\
\midrule
BM25        & 14.4 & 16.0 & 13.4 & 19.0 & \cellcolor{green!15}24.4 & \cellcolor{green!15}27.8 & \cellcolor{green!20}31.3 & \cellcolor{green!25}42.1 \\
SBERT       & 12.8 & 14.6 & 11.4 & 17.1 & \cellcolor{green!20}29.9 & \cellcolor{green!20}32.8 & \cellcolor{green!30}38.0 & \cellcolor{green!30}46.0 \\
BGE         & 12.7 & 14.0 & 11.6 & 16.1 & \cellcolor{green!15}25.7 & \cellcolor{green!18}28.6 & \cellcolor{green!25}33.3 & \cellcolor{green!30}41.6 \\
Contriever  & 10.2 & 11.0 &  8.7 & 12.6 & \cellcolor{green!18}25.7 & \cellcolor{green!20}28.7 & \cellcolor{green!25}32.3 & \cellcolor{green!30}41.3 \\
Nomic       & 13.4 & 14.9 & 12.0 & 17.4 & \cellcolor{green!5}16.6  & \cellcolor{green!5}18.1  & \cellcolor{green!10}20.5 & \cellcolor{green!10}24.8 \\
Inst-L      & 12.5 & 14.1 & 12.9 & 18.0 & \cellcolor{green!25}32.2 & \cellcolor{green!25}36.5 & \cellcolor{green!30}41.2 & \cellcolor{green!40}54.2 \\
Inst-XL     & 19.1 & 21.2 & 18.6 & 25.2 & \cellcolor{green!18}34.5 & \cellcolor{green!20}38.0 & \cellcolor{green!30}44.2 & \cellcolor{green!35}55.2 \\
E5-Mistral  & 14.3 & 15.8 & 13.5 & 18.6 & \cellcolor{green!30}40.1 & \cellcolor{green!30}44.2 & \cellcolor{green!40}52.8 & \cellcolor{green!45}64.6 \\
SFR-Mistral & 15.0 & 16.8 & 14.6 & 20.9 & \cellcolor{green!30}42.8 & \cellcolor{green!35}46.7 & \cellcolor{green!40}55.8 & \cellcolor{green!45}67.6 \\
GTE-Qwen    & 19.3 & 21.2 & 16.9 & 24.0 & \cellcolor{green!28}43.1 & \cellcolor{green!30}47.9 & \cellcolor{green!40}54.6 & \cellcolor{green!45}68.2 \\
GTE-Qwen2   & 18.8 & 20.7 & 18.3 & 24.7 & \cellcolor{green!30}43.9 & \cellcolor{green!30}47.5 & \cellcolor{green!40}56.5 & \cellcolor{green!45}67.2 \\
ReasonIR    & 20.3 & 22.5 & 18.5 & 26.3 & \cellcolor{green!22}38.6 & \cellcolor{green!25}42.8 & \cellcolor{green!35}51.7 & \cellcolor{green!40}63.7 \\
RaDeR       & 17.8 & 19.7 & 17.0 & 23.5 & \cellcolor{green!28}41.8 & \cellcolor{green!30}46.0 & \cellcolor{green!40}53.8 & \cellcolor{green!45}66.0 \\
Diver       & 26.1 & 28.4 & 23.7 & 32.7 & \cellcolor{green!22}43.8 & \cellcolor{green!25}48.2 & \cellcolor{green!35}56.3 & \cellcolor{green!40}69.6 \\
\bottomrule
\end{tabular}}
\label{tab:long_context}
\end{table}



\subsection{Query Perturbation Robustness}

\begin{figure}[h]
\centering
\includegraphics[width=0.8\columnwidth]{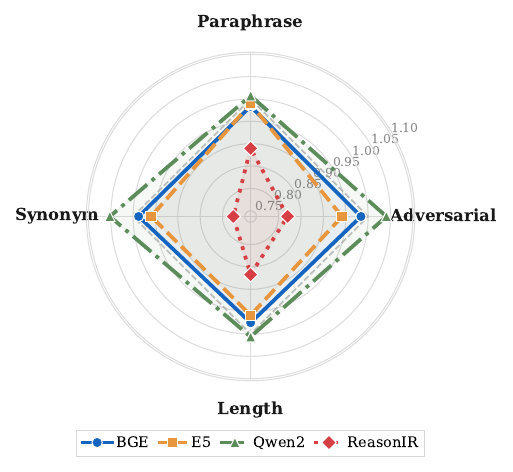}
\caption{Robustness under query perturbations, shown as retention ratio: $\text{nDCG@10}_{\text{perturbed}} / \text{nDCG@10}_{\text{original}}$. Values above 1.0 indicate improvement; below 1.0 indicate degradation.}
\label{fig:robustness}
\end{figure}

We evaluate four representative retrievers---BGE (small dense), E5-Mistral (LLM-based dense), GTE-Qwen2 (LLM-based, strongest general-purpose), and ReasonIR (reasoning-specialized)---under four perturbation types applied to all 12 BRIGHT tasks.
\emph{Paraphrasing} generates multiple lexically distinct rephrasings of each query while preserving its semantic intent.
\emph{Synonym replacement} substitutes content words with WordNet synonyms at increasing intensity levels (1--3 replacements per query).
\emph{Adversarial token insertion} injects semantically unrelated distractor tokens into the query at varying densities (1--2 tokens).
\emph{Length perturbations} either expand the query by appending contextual elaborations or contract it by removing non-essential tokens.
Figure~\ref{fig:robustness} reports the retention ratio (perturbed nDCG@10 divided by original nDCG@10) averaged across all 12 tasks. The four models exhibit strikingly different robustness profiles.
GTE-Qwen2 is the most stable and actually \emph{improves} under several perturbation types: synonym replacement yields a retention of 1.05 (24.5 vs.\ 23.3 original), and adversarial insertion yields 1.04 (24.3 vs.\ 23.3). This suggests that Qwen2's large-scale pretraining provides sufficient lexical diversity to absorb token-level noise, and that certain perturbations may inadvertently add useful query terms.
BGE is similarly robust (all retentions $\approx$~0.98--0.99), consistent with its compact architecture being less sensitive to individual token changes.

In contrast, ReasonIR is substantially more brittle: synonym replacement drops performance from 24.1 to 18.7 (retention 0.78; $-$22.4\%), and adversarial insertions reduce it to 19.8 (retention 0.82; $-$17.8\%).
This fragility is concerning because reasoning-specialized retrievers are designed for complex queries that are precisely the type most likely to undergo reformulation in practice.
The vulnerability likely arises from ReasonIR's training on carefully structured reasoning queries, making it sensitive to deviations from expected token patterns.
E5-Mistral shows moderate degradation (17.9 to 16.9 under adversarial; retention 0.94), placing it between the robust generalists and the fragile specialist.

Overall, robustness gaps are driven primarily by lexical-level noise (synonyms, adversarial tokens) rather than paraphrasing, suggesting that token-distribution shifts pose a greater risk than semantic reformulation.

\subsection{Hybrid Fusion Strategies}

\begin{figure}[t]
\centering
\includegraphics[width=\columnwidth]{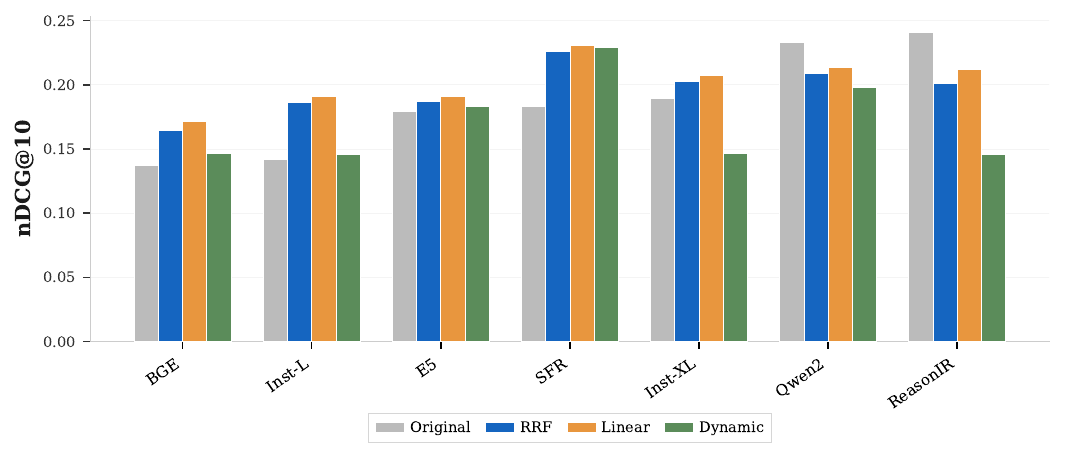}
\caption{Hybrid fusion of BM25 with seven dense retrievers (12-task average nDCG@10). Three strategies shown: Reciprocal Rank Fusion (RRF), Linear score combination ($\alpha{=}0.5$), and Dynamic Weight Adaptation (DAT).}
\label{fig:fusion}
\end{figure}

We evaluate now whether combining BM25's sparse lexical signals with dense retrievers can improve the results achieved by either component alone.
We fuse BM25 with seven dense retrievers using three strategies: Reciprocal Rank Fusion (RRF; $k{=}60$)~\cite{cormack2009reciprocal}, Linear score combination (min--max normalized, $\alpha{=}0.5$), and Dynamic weight Adaptation (DAT)~\cite{hsu2025dat}, which adjusts the interpolation weight $\alpha$ per query based on the score distribution overlap between the sparse and dense result lists---assigning higher weight to the component whose scores show greater separation between its top-ranked and lower-ranked documents.
Figure~\ref{fig:fusion} reports the 12-task average nDCG@10 for each combination. Linear Fusion is the most consistently effective strategy for mid-tier dense retrievers.
It improves BGE from 13.7 to 17.2 (+3.5 nDCG@10), Instructor-L from 14.2 to 19.1 (+4.9), and SFR-Mistral from 18.3 to 23.1 (+4.8)---gains that rival those of reasoning augmentation (\S\ref{sec:efficiency}) but require no LLM inference at query time.
RRF follows similar trends but produces smaller improvements, consistent with its rank-only aggregation discarding magnitude information.
Inst-XL also benefits (+1.8 under Linear), and E5-Mistral shows a modest gain (+1.2).

However, fusion can \emph{harm} already-strong retrievers.
GTE-Qwen2 drops from 23.3 to 21.3 under Linear ($-$2.0), and ReasonIR drops from 24.1 to 21.2 ($-$2.9).
These regressions occur because BM25's lexical signal conflicts with the dense retriever's semantic ranking when the dense model is already effective: the fused list promotes lexically similar but semantically irrelevant documents.
The Dynamic (DAT) strategy is unstable, producing large regressions for several models (e.g., ReasonIR drops to 14.6); we include it for completeness but do not recommend it without further calibration. The practical recommendation is straightforward: Linear Fusion with BM25 is a reliable, zero-cost upgrade for any dense retriever scoring below $\sim$20 nDCG@10 on BRIGHT, but should be avoided for top-tier models where it introduces more noise than signal.

\begin{table}[h]
\centering
\caption{Query performance prediction via confidence calibration (AUROC), using the top-ranked retrieval score as a predictor of query-level success (gold document in top-$k$). This setup follows the query performance prediction (QPP) paradigm~\cite{carmel2010estimating}. Quality thresholds: \textcolor{orange!80!black}{Fair} (0.55--0.65), \textcolor{red!70!black}{Poor} ($<$0.55). 
}

\label{tab:calibration}
\resizebox{0.8\columnwidth}{!}{%
\begin{tabular}{ll cccc c}
\toprule
\textbf{Category} & \textbf{Model} & \textbf{@5} & \textbf{@10} & \textbf{@25} & \textbf{@50} & \textbf{Quality} \\
\midrule
Sparse & BM25 & 0.597 & \textbf{0.602} & \textbf{0.611} & \textbf{0.604} & \textcolor{orange!80!black}{Fair} \\
\midrule
\multirow{4}{*}{Dense} & SBERT & 0.536 & 0.544 & 0.525 & 0.519 & \textcolor{red!70!black}{Poor} \\
 & Contriever & 0.565 & 0.575 & 0.560 & 0.526 & \textcolor{orange!80!black}{Fair} \\
 & Nomic & 0.587 & 0.580 & 0.581 & 0.570 & \textcolor{orange!80!black}{Fair} \\
 & BGE & 0.541 & 0.538 & 0.528 & 0.519 & \textcolor{red!70!black}{Poor} \\
\midrule
\multirow{3}{*}{Instruct} & Inst-L & 0.508 & 0.528 & 0.528 & 0.521 & \textcolor{red!70!black}{Poor} \\
 & Inst-XL & 0.571 & 0.564 & 0.546 & 0.540 & \textcolor{orange!80!black}{Fair} \\
 & SFR-Mistral & 0.568 & 0.550 & 0.553 & 0.551 & \textcolor{orange!80!black}{Fair} \\
\midrule
\multirow{3}{*}{LLM-Based} & E5-Mistral & 0.537 & 0.546 & 0.552 & 0.544 & \textcolor{red!70!black}{Poor} \\
 & GTE-Qwen & \textbf{0.598} & 0.594 & 0.571 & 0.583 & \textcolor{orange!80!black}{Fair} \\
 & GTE-Qwen2 & 0.564 & 0.570 & 0.561 & 0.552 & \textcolor{orange!80!black}{Fair} \\
\midrule
\multirow{3}{*}{Reasoning} & Diver & 0.577 & 0.584 & 0.554 & 0.549 & \textcolor{orange!80!black}{Fair} \\
 & RaDeR & 0.568 & 0.550 & 0.560 & 0.546 & \textcolor{orange!80!black}{Fair} \\
 & ReasonIR & 0.556 & 0.555 & 0.552 & 0.536 & \textcolor{orange!80!black}{Fair} \\
\bottomrule
\end{tabular}%
}
\end{table}

\subsection{Query Success Prediction}

Finally, we assess whether retrieval scores can serve as a simple and reliable signal for query-level success prediction---a task closely related to query performance prediction (QPP)~\cite{carmel2010estimating}, which aims to estimate retrieval quality without relevance judgments. We adopt a post-retrieval QPP setup, using the top-1 retrieval score as a predictor of query-level success. While not a formal probabilistic calibration, AUROC provides a threshold-independent measure of how well the retrieval scores discriminate between successful and failed queries, which is critical for confidence-aware routing, selective abstention, and downstream RAG pipelines.

For each query, we label it successful if a gold document appears within the top-$k$, and
compute AUROC@$\{5,10,25,50\}$ across all 12 BRIGHT tasks (Table~\ref{tab:calibration}).  Confidence calibration is uniformly weak across all model families with significant overlap between the scores of successful and failed queries.
The best-calibrated model is BM25, which achieves AUROC@10 = 0.602 and AUROC@25 = 0.611 only marginally above chance (0.50).
GTE-Qwen is competitive at strict cutoffs (AUROC@5 = 0.598) but degrades at deeper ranks.
Dense bi-encoders are consistently poorly calibrated: SBERT (0.544), BGE (0.538), and E5-Mistral (0.546) at @10 are barely above random.
Reasoning-specialized models perform only slightly better (Diver: 0.584, RaDeR: 0.550, ReasonIR: 0.555 at @10), despite their superior effectiveness.

The gap between effectiveness and confidence is notable.
Diver achieves 29.4 nDCG@10 (best overall) but only 0.584 AUROC@10; BM25 at 14.5 nDCG@10 provides better confidence separation (0.602). This indicates that retrieval scores encode relevance \emph{ranking} information but not reliable \emph{absolute} confidence.
All ROC curves cluster near the diagonal, indicating that no model reaches the reliability threshold needed for autonomous confidence-based routing without additional calibration mechanisms such as score normalization, ensemble agreement, or learned confidence estimators.
\section{Discussion}
\label{sec:discussion}
Our evaluation reveals several cross-cutting insights with direct practical implications.

\paragraph{Architectural efficiency matters more than scale.}
Diver (4B parameters) Pareto-dominates all 7B LLM-based dense models on both effectiveness and throughput (29.4 nDCG@10 at 47.3 QPS vs.\ GTE-Qwen2's 23.3 at 6.0 QPS), suggesting that targeted training on reasoning-intensive data yields better returns per FLOP than simply increasing model size.
Practitioners should prioritize retrieval-specialized architectures over generic LLM bi-encoders for reasoning-intensive tasks.

\paragraph{Reasoning augmentation requires task-aware routing.}
Chain-of-thought augmentation produces large gains on science domains with complex, implicit information needs (Biology: +15 nDCG@10) but consistently \emph{degrades} performance on formal domains like AoPS and LeetCode where queries are in the form of already precise specifications.
This argues against blanket deployment; a lightweight domain classifier should gate whether augmentation is applied.

\paragraph{Robustness--effectiveness is a hidden trade-off.}
ReasonIR achieves 24.1 nDCG@10 but loses up to 22\% under synonym replacement---a perturbation routine in real queries---while GTE-Qwen2 is robust but less effective.
This trade-off is invisible in standard benchmarks and represents a deployment risk.

\paragraph{Confidence calibration remains unsolved.}
No model family produces well-calibrated confidence scores (best AUROC $\approx$ 0.60), with the weakness consistent across scales and training paradigms.
This limits the utility of retrieval scores for downstream routing in RAG pipelines without additional calibration mechanisms.

\section{Conclusion}
\label{sec:conclusion}
We presented a comprehensive reproducibility study of the BRIGHT benchmark, extending evaluation to efficiency, robustness, and confidence calibration across 14 retrievers and 12 tasks. We make five take-home messages:
\textbf{(1)}~Reasoning-specialized retrievers Pareto-dominate larger general-purpose LLM encoders---scale alone does not justify cost.
\textbf{(2)}~Reasoning augmentation is free for sub-1B models but should be applied selectively by domain.
\textbf{(3)}~BM25 linear fusion reliably improves any retriever below $\sim$20 nDCG@10 at zero inference cost, but harms top-tier models.
\textbf{(4)}~Robustness varies dramatically and is not predicted by effectiveness; perturbation testing should become standard.
\textbf{(5)}~Raw retrieval scores are unreliable for confidence-based routing across all families; additional calibration is needed.
We release all code and artifacts for reproducibility.

\bibliographystyle{ACM-Reference-Format}
\bibliography{sample-base}

\appendix

\end{document}